\documentclass[preprintnumbers, amssymb,aps,pra,twocolumn]{revtex4}
\usepackage{amsmath}
\usepackage{graphicx}
\usepackage{color}

\usepackage[usenames,dvipsnames]{xcolor}
\usepackage{siunitx}
\usepackage[T1]{fontenc}
\usepackage[utf8]{inputenc}
\usepackage{ulem}
\usepackage{subfigure}
\usepackage[colorlinks=true, letterpaper=true, pdfstartview=FitV, linkcolor=red, citecolor=blue, urlcolor=red]{hyperref}

\begin{document}

\title{Triangular vortex lattices and giant vortices in rotating bubble Bose--Einstein condensates}

\author{Angela C. White}
\email{ang.c.white@gmail.com}
\affiliation{%
 Australian Research Council Centre of Excellence in Future Low-Energy Electronics Technologies, School of Mathematics and Physics, University of Queensland, St. Lucia, Queensland 4072, Australia
}
\affiliation{%
Australian Research Council Centre of Excellence for Engineered Quantum Systems, School of Mathematics and Physics, University of Queensland, St. Lucia, Queensland 4072, Australia
}

\begin{abstract} 
 We show that three-dimensional spherical-shell condensates respond to rotation by forming two aligned triangular Abrikosov-like vortex lattices on each hemispherical surface. The centrifugal force due to rotation causes an elliptical deformation of the spherical shell condensate shape and for faster rotation rates, drives the formation of a central multi-charged vortex-antivortex pair at the poles surrounded by a ring of singly charged vortices in the bulk density. The vortex distributions observed in each hemisphere take a similar form to those found in rotating harmonic plus quartic traps. 
 
\end{abstract}
\maketitle

\newpage

Bose--Einstein condensates (BECs) are versatile and highly tuneable quantum systems that have been created in a variety of simply and multiply connected topologies, from disks, cigars, and spheres, to toroids. BECs have been intensely studied for their fundamental physics properties such as superfluidity \cite{Madison2000,Ketterle2001,Phillips2011} and recently increasing attention has been focused on their great promise in technological applications in areas including precision inertial sensing \cite{Bongsreview} and atomtronics \cite{atomtronics}. Although creating spatially dependent dressed states by radio-frequency coupling was first proposed to engineer a shell geometry a decade ago \cite{Garroway}, achieving superfluid condensate `bubbles' has been prohibitively difficult in a terrestrial environment. The difficulty arises due to the need to compensate for gravity \cite{Perrin2004,Dubessy2022}, which otherwise causes the condensate to sag into a hemispherical bowl. The realisation of Bose--Einstein condensates (BECs) confined to a thin hollow shell in the micro-gravity environment of the Cold Atomic Laboratory aboard the International Space Station \cite{LundbladNature} circumvents this problem and has opened up the study of this newly accessible simply-connected topology \cite{Perspective}. Recently, optically trapped immiscible Bose--Bose mixture experiments in a terrestrial environment have been shown to provide an alternative avenue for realising shell geometries \cite{Jia2022}, with  the creation of a shell condensate of one species surrounding an inner spherically-shaped condensate core of a different atomic species \cite{Wolf2022}. Such ball and shell structures were also demonstrated in early terrestrial experiments with magnetically trapped hyperfine states of $^{87}$Rb \cite{Hall1998,Mertes2007}. The sizes of shells achievable in terrestrial mixture experiments are smaller than those attainable in a micro-gravity environment.  These experimental prospects have prompted theoretical investigations into the unique properties of shell-shaped condensates, including the critical temperature of Bose--Einstein condensation \cite{SalasnichPRL2019,SalasnichPRL2020,Vishveshwara2021}, collective excitation dynamics \cite{Lannert2018} and finite-size Berezinski--Kosterlitz--Thouless (BKT) physics \cite{SalasnichPhysRevR2022}.  The closed curved surface of the bubble geometry promises interesting responses to rotation and motivates further study into vortex dynamics on curved surfaces \cite{NelsonReview,Bowen2019,Fetter2021sphericalfilm,Fetter2022Ellipsoid}. 

Classical fluid flow on the surface of spheres has long been of interest due to its relevance in describing planetary atmospheric dynamics \cite{Jovian1994,Verdiere94}. Studies have shown notable differences in vortex dynamics  in comparison to planar geometries, as the curvature of the surface leads to weaker interaction between different parts of the flow \cite{Dritschel1992}. Superfluid flow on the surface of spheres similarly promises intriguing physics arising from the curvature and topology of a spherical shell as a simply connected surface. In superfluids, the superfluid velocity is a continuous field {$\textbf{v}= \hbar \nabla \theta/m$}, purely dependent on the gradient of the condensate phase $\theta$. Here $\hbar$ is the reduced Planck constant and $m$ the atomic mass. For superflow in bubble condensates, a consequence of the continuous nature of the superfluid velocity field is that single vortices that begin on the outer surface of the bubble and end on the inner surface of the bubble cannot exist in isolation. A second vortex or topological defect must be present to heal the velocity field. This requirement arises from a special case of the Poincar{\'e}--Hopf theorem, known as the hairy-ball or Hedgehog theorem on a spherical-shell, and has been established for classical point-vortex systems \cite{Lamb}.  In superfluid bubbles, long range attractive interactions between vortex-antivortex pairs result in their relaxation towards the equator and eventual annihilation being energetically preferred \cite{Padavic2020}. However, in the presence of an external rotation, the minimum energy configuration consists of vortex pairs, with the second vortex rotating with opposite circulation aligned at the antipode of the condensate shell \cite{Lamb}. The critical rotation velocity to stabilise vortex-antivortex pairs at the poles and its dependence on the dimensionality and thickness of the superfluid bubble has been explored in detail in \cite{Padavic2020}. In particular, the stability of vortices was found to be a distinguishing feature of shell geometries in comparison to their filled counterpart, which may be a useful experimental indicator of the underlying topology of the condensate \cite{Padavic2020}. 

 \begin{figure*}[!tb]
\centering
\includegraphics[width=1.6\columnwidth]{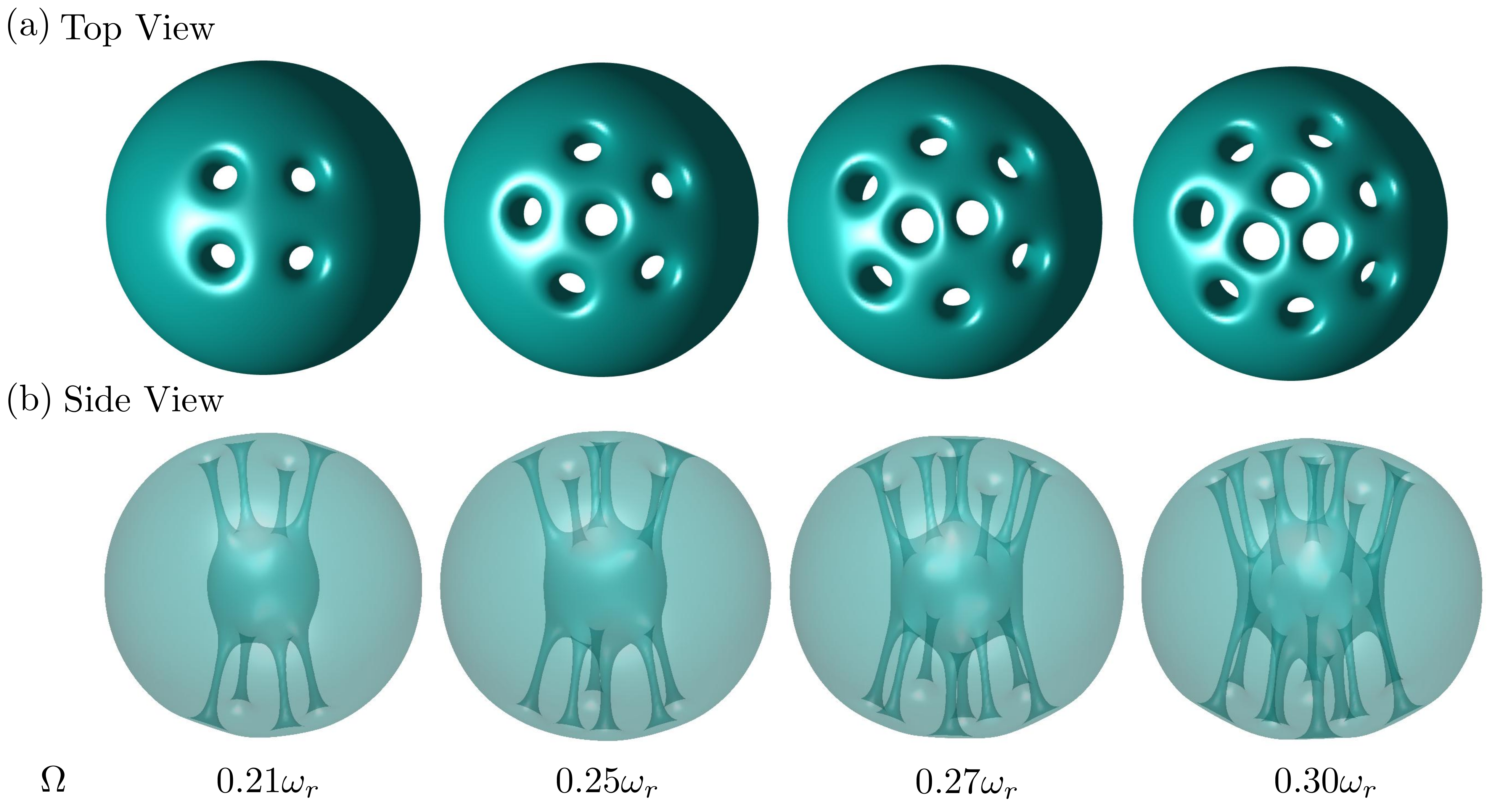}
\caption{Isosurfaces of constant density of shell-shaped Bose--Einstein condensates under rotation, with rotation frequencies from left to right of $0.21\omega_{r}$, $0.25\omega_{r}$, $0.27\omega_{r}$, and $0.3\omega_{r}$. The minimum energy states show the numbers of vortices increase as the external rotation frequency is increased. a) Top row: top view of holes in the isosurface (plotted at $50\%$ of the maximum condensate density) show the position of vortex cores. b) Bottom row: side view of lines in the isosurface (plotted at $10\%$ of the maximum condensate density) show the line length of vortex cores ending on an inner central sphere, which corresponds to the inner surface of the shell.  }
\label{figure1}
\end{figure*}
 
Condensates confined to filled spheres and disks respond to an externally imposed rotation above a critical velocity by the formation of a triangular so-called Abrikosov lattice of vortices. Triangular lattice formation in these geometries is one of the hallmark's of superfluidity and has been verified in early BEC experiments \cite{Madison2000,Ketterle2001,KetterlePRL2003} and extensively studied theoretically \cite{BaymPRL2003,Fetter2009,LeePRA2016,Lee2PRA2016}. In bubble condensates, there has been little investigation into the response to rotation beyond point-vortex models. A study employing point-vortex model solutions as a variational ansatz of the Gross--Pitaevskii equation suggested rotating solutions of a ring of equally spaced positive (and negative) vortices on each hemisphere \cite{Gelantalis2012}. Two vortex configurations were found, one in which each vortex is aligned with its anti-vortex pair, and another skewed solution in which the vortex and anti-vortex rings are misaligned, with each anti-vortex on one ring positioned half-way between each vortex on the other ring \cite{Gelantalis2012}. Such vortex ring solutions contrast to the typical Abrikosov vortex lattices observed in pancake geometries, raising the question of if triangular lattices also occur in superfluid shell structures and which configuration is naturally occurring for bubble condensates under rotation. A recent preprint we became aware of while preparing this manuscript focused on studying fluid excitations known as Rossby waves occurring on two-dimensional shells, forced approximate vortex lattice solutions by applying an additional potential compensating for the centrifugal force \cite{Saito}.  \cite{Efimkin} also apply vortex-lattice solutions in two dimensional condensate shells, in order to investigate equatorial waves. 

In this article we establish the response of a three-dimensional bubble condensate to an external rotation, going beyond point-vortex models and variational calculations and numerically solve the mean-field Gross--Pitaevskii equation.  We find at slow rotation rates, aligned triangular Abrikosov vortex lattices develop in each hemispherical shell. As the rotation rate is increased, we see a transition to a multi-charged vortex at each pole, surrounded by singly charged vortices in the bulk. The vortex distribution in each hemisphere is the same as that observed in disk-shaped condensates confined in harmonic plus quartic potentials \cite{Fetter2005}. This similarity can be explained by the local form of the shifted harmonic shell potential at the poles, with quartic and higher order terms dominating the local effective trapping potential in these regions. Additionally, we observe a distortion of the spherical-shell shape as the rotation rate is increased due to the centrifugal barrier which causes atoms to move away from the poles, and so the condensate takes on an elliptical shape with greater density concentrated around the equator. Finally we discuss what this may imply for reaching the quantum-hall regime in bubble geometries under rotation and future challenges in experimentally verifying these findings.    

We model a Bose--Einstein condensate in the zero temperature limit by solving the three-dimensional Gross--Pitaevskii equation for the mean-field condensate wave-function $\psi$.  Under an external imposed rotation around the $z$ axis, the Gross--Pitaevskii equation in the co-rotating frame takes the form
\begin{equation} \label{GPE}
i \hbar \frac{\partial}{\partial t} \psi = \left( -\frac{\hbar^2}{2m}\nabla^{2}+V+Ng\left|  \psi \right|^2 -\Omega_{z}L_{z} \right)\psi.
\end{equation}
The condensate bubble is confined to a shifted harmonic potential $V=m\omega_{r}^2(r-r_{0})^{2}/2$, with a harmonic trapping frequency $\omega_{r}=15.9\times 2\pi$ Hz shifted by the central shell radius $r_{0}=15$ $\mu \text{m}=5.5a_{\text{osc}}$ where $a_{\text{osc}}=\sqrt{\hbar/m\omega_{r}}$ is the harmonic oscillator length.  Here $r^2=x^2+y^2+z^2$. The interaction between bosons in the condensate is described by $g=4\pi\hbar^{2}a/m$ in terms of the three-dimensional scattering length $a$. We numerically model a condensate of $N=2\times10^5$ Rb$^{87} $ atoms with an s-wave scattering length of $98a_{0}$. The condensate wavefunction is normalised to unity, $\int |\psi|^{2}\text{d}\textbf{x}^3=1$.
The rotation frequency around the $z$ axis is given by $\Omega_{z}$, and $L_{z}=xp_{y}-yp_{x}$ 
is the angular momentum operator. In the absence of rotation, the condensate density drops to below $5\%$ of the maximum density at radii smaller than $2.5a_{\text{osc}}$ and greater than $8.37a_{\text{osc}}$. 
To find the minimum energy states for a particular rotation frequency, we solve Equation \ref{GPE} using imaginary time propagation, replacing $t\rightarrow -i \tau$ \cite{Chiofalo}. We employ a split-step method \cite{Javanainen2006}, scaling energy in units of $\hbar \omega_{r}$ and lengths in units of harmonic oscillator length $a_{\text{osc}}$.  Equation \ref{GPE} is modelled on a three-dimensional grid of $256^3$ points over a spatial extent of $(16 a_{\text{osc}})^3$ for small external rotation frequencies, increasing up to $ (20 a_{\text{osc}})^3$, as the rotation frequency and consequently the radius of the condensate bubble increases. 

We first investigate the regime of small external rotation frequencies, when the rotation frequency is less than a critical frequency $\Omega<\Omega_{c}$. 
In this regime, we find the condensate responds to rotation by creating a triangular lattice of vortices in each hemisphere. Each vortex in the lattice begins on the outer edge of the bubble, extending through to and ending on the curved central surface of the shell. The co-rotating vortices in the top hemisphere of the bubble are mirrored by an aligned lattice of anti-rotating vortices in the bottom hemisphere. The direction of vortex rotation is defined as the direction of rotation as seen from the outer surface of the shell. An example of these lattice configurations are depicted in figure \ref{figure1}. The vortices form a triangular lattice in each hemisphere of the shell, exhibiting qualitatively the same Abrikosov-like distribution as the triangular vortex lattices observed in harmonically trapped condensates.  

 \begin{figure}[!tbh]
\centering
\includegraphics[width=\columnwidth]{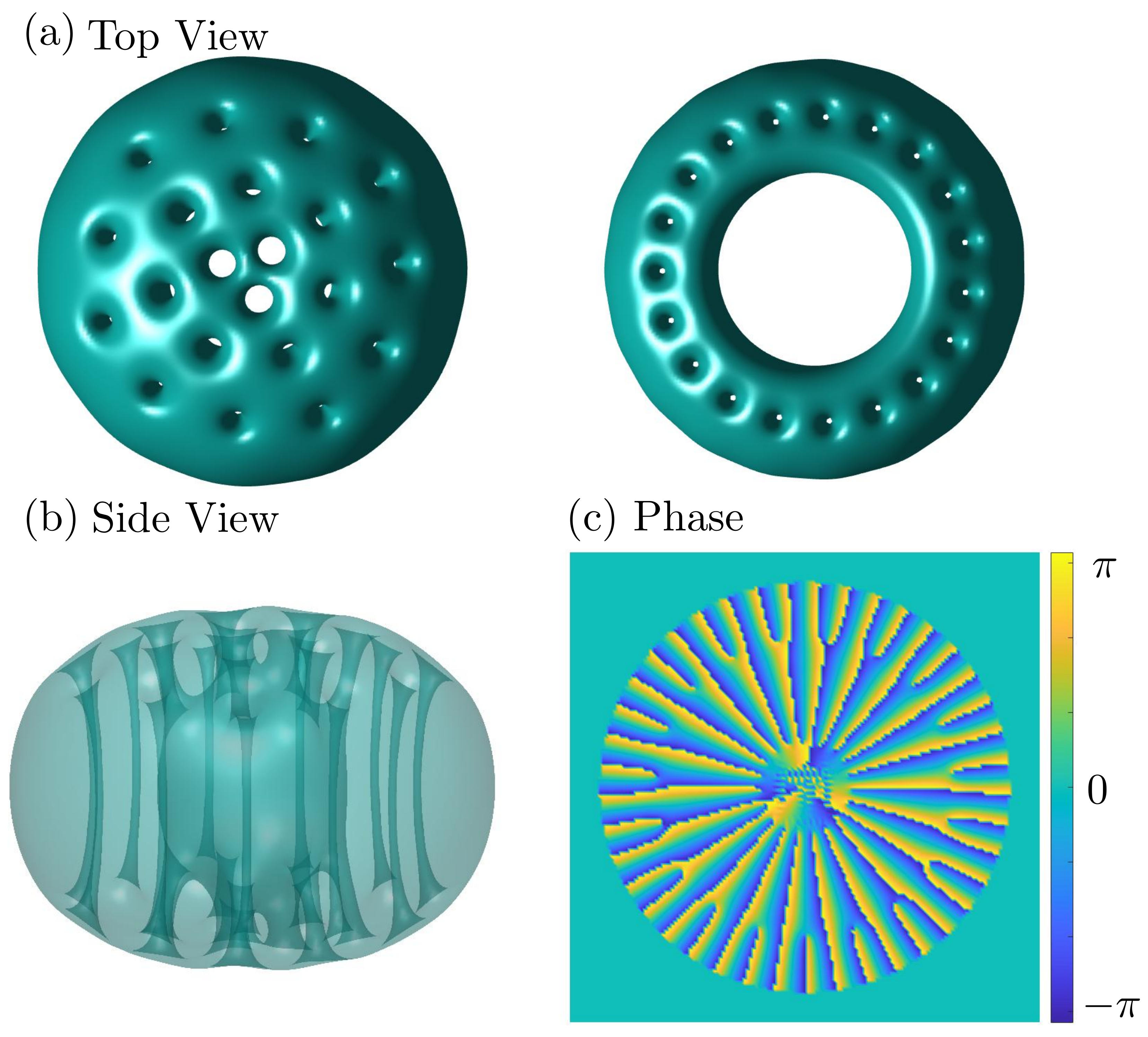}
\caption{ a) Top view of isosurfaces plotted at $10\%$  of the maximum condensate density for shell-shaped Bose--Einstein condensates under a rotation of $0.4 \omega_{r}$ (left column) and  $0.54\omega_{r}$ (right column). b) The corresponding side view of the constant density isosurface for an external rotation of $0.4 \omega_{r}$, showing vortex lines through the bulk condensate density and an elliptical shaped density distribution. c) The phase profile of a condensate experiencing an external rotation frequency of  $0.54\omega_{r}$ at $\theta(x,y,0)$  illustrates the giant multi-charged vortex of winding 22, that has formed at the poles. This multi-charged vortex is surrounded by 19 singly charged vortices in the bulk condensate density. }
\label{figure2}
\end{figure}

As the rate of external rotation is increased, there is a transition to vortices that do not end on the inner surface of the shell but begin and end in opposite hemispheres on the outer surface of the shell (see figure \ref{figure2} (b)). Such vortex lines are known as U-shaped vortices, with the name originally coined as their shape resembles a wide `U' \cite{Rosenbusch2002}. These U-shaped vortex lines traverse the bulk condensate density and are longer in length (and therefore cost more energy) than vortices that end on the inner surface of the shell. The appearance of U-shaped vortices coincides with a visible distortion of the spherical shell shape that occurs as a consequence of the centrifugal force, which takes the form $-m\Omega^2 r^2/2$, and pushes atoms away from the center of the condensate. As a result, the thickness of the condensate density varies around the shell, with thicker regions closer to the equator and a thinner condensate width at the poles. An example of such an asymmetric density distribution can be seen in figure \ref{figure2} (b).  
The unequal density distribution around the shell is also reflected in the size of the vortex cores (see figure \ref{figure2} (a) column 1 and (b)). The vortex cores closest to the poles have a larger core size than those closer to the equator, as the local coherence length ($ \xi = \sqrt{\hbar^2/(2mng)}$ \cite{PethickSmith} where $n=|\psi|^2$ is the local background condensate density) becomes smaller closer to the equator, where the background condensate density is greater. This distribution contrasts to vortices in harmonically trapped condensates,  which have larger core sizes at the edge of the harmonic trap and thinner cores at the center of the condensate where the background condensate density is larger. Deformation of a condensate due to the centrifugal force has been observed for a superfluid under fast rotation and confined in a shell trap \cite{Perrin2020}. 

The third distinctive regime we observe with a further increase in the external rotation rate is marked by the transition from a shell to a toroidal geometry, with the appearance of a central multi-charged vortex core of opposite winding at each pole. An example of typical density configurations in this regime is given for an external rotation frequency of $0.54\omega_{r}$, shown in figure \ref{figure2} (a) column 2. Here, a ring of singly charged vortices is observed surrounding a central hole that is formed by the core of a multi-charged vortex of winding 22. The corresponding slice of the phase profile in the $x$-$y$ plane for $z=0$ is depicted in figure \ref{figure2} (c), where each phase slip in the outer ring (of phase slips) indicate a singly charged vortex and the inner ring of phase slips correspond to the circulation of the multi-charged vortex. This giant-vortex regime that occurs at faster angular velocities is reminiscent of giant vortices observed in rapidly rotating condensates confined in harmonic-plus quartic traps \cite{Lundh2002, Kasamatsu2002, Fetter2005, Zaremba2006} and indeed suggests that a single pair of multi-charged vortices at the poles may arise as the external rotation rate is further increased. Giant vortices are a feature of fast rotation of condensates in trapping potentials that are steeper than harmonic, and consequently also develop in condensates trapped in rotating hard-walled buckets \cite{Fischer2003}. 

The observation of similar vortex configurations in both harmonically shifted bubble potentials and rotating condensates confined in harmonic plus quartic traps can be explained by the locally quartic nature of the bubble potential around the minimum in the trapping potential. This can be seen by writing the trapping potential in cartesian coordinates and applying a Taylor expansion around the trap minimum. The leading terms in the Taylor expansion around the trap minimum are locally quartic: $V(x,0,r_{0}) \propto x^4$ (and similarly $V(r_{0},y,0)\propto y^4$ and $V(0,r_{0},z)\propto z^4$). This locally quartic behaviour goes some way to establishing the likeness between the response to rotation of spherical shells in comparison to harmonic plus quartic and other steeply confined BECs. Note however, the geometries are not entirely equivalent as due to the shape of the shell trapping geometry, atoms can move from the quartic - like region to lower $z$. Indeed atoms are pushed out by the centrifugal barrier. 

In addition to suggesting that a transition to a single macroscopic giant vortex pair located at the poles may occur, the locally quartic behaviour also implies that quantum hall states may be difficult to reach for shell condensates. In harmonically trapped condensates, as the external rotation rate approaches the trapping frequency ($\Omega \approx\omega_{r}$), and the centrifugal potential exactly cancels out the trap potential, the regime of the lowest-Landau-level approximation, or quantum Hall regime, is reached  \cite{Ho2001} and the lowest-energy state is the Bosonic Laughlin state \cite{Laughlin1983}.  Early experiments have explored condensate physics approaching the lowest Landau level regime, with BECs rotating close to the trap frequency \cite{Cornell2004,Dalibard2004}. Recently geometric squeezing was applied to prepare condensates directly in the lowest Landau level \cite{Zwierlein2021} and the evolution of a interacting BEC occupying a single Landau gauge wavefunction has been subsequently demonstrated \cite{Zwierlein2022}. While exploring the effect of geometry and topology on Landau-level physics in a bubble geometry would potentially uncover interesting physics, the locally quartic nature of shell condensates suggests this regime may be out of reach. Previous work indicates that the locally quartic nature of the trap prohibits attaining the Bosonic Laughlin states at least experimentally, as an additional weak quartic potential renders them highly fragile \cite{vKlitz2019}. In the rapidly rotating regime for a weakly anharmonic trapping potential, the Laughlin state is fragile and energetically unfavourable, and the energetically favourable state is a giant-vortex state \cite{vKlitz2019}.  

In conclusion, we have demonstrated that the response to rotation of three dimensional bubble condensates can be classified into three distinct regimes. The first regime, for slow external rotation rates is characterised by the formation of two aligned triangular Abrikosov-like vortex lattices in each hemisphere, with all constituent vortices beginning on the outer bubble surface and ending on the inner surface. As the external rotation rate is increased, the centrifugal barrier results in a distortion of the spherical shell shape and we enter a regime where U-shaped vortex lines that traverse the bulk condensate density are formed closer to the equator. Finally, we find that at faster rotation rates, a giant vortex--antivortex pair forms at the poles, surrounded by singly charged vortices in the bulk condensate density. This regime corresponds to a transition of topology from a spherical-shell shape to a toroid. Creating spherical-shaped shells experimentally is currently challenging and to date shells of cold atoms created in the microgravity environment of the international space station are elliptical in shape, with non-uniform width \cite{LundbladNature}. We leave investigating the effect of these changes in geometry on the resulting vortex lattice structures created under rotation as an avenue for future work. 

{\bf Acknowledgments:}
I acknowledge and thank A.~Benseny, J.~J.~Hope, A.~Fetter and W. von Klitzing for useful discussions and enlightening comments. This research was undertaken with computational resources provided by the National Computational Infrastructure (NCI), which is supported by the Australian Government. This research was supported by the Australian Research Council Centre of Excellence in Future Low-Energy Electronics Technologies (project number CE170100039) and the Australian Research Council (ARC) Centre of Excellence for Engineered Quantum Systems (EQUS, CE170100009). 
\bibliographystyle{apsrev4-1}

\end{document}